# A super-Earth transiting a nearby low-mass star

David Charbonneau[1], Zachory K. Berta[1], Jonathan Irwin[1], Christopher J. Burke[1], Philip Nutzman[1], Lars A. Buchhave[1,2], Christophe Lovis[3], Xavier Bonfils[3,4], David W. Latham[1], Stéphane Udry[3], Ruth A. Murray-Clay[1], Matthew J. Holman[1], Emilio E. Falco[1], Joshua N. Winn[5], Didier Queloz[3], Francesco Pepe[3], Michel Mayor[3], Xavier Delfosse[4] & Thierry Forveille[4]

[1]Harvard-Smithsonian Center for Astrophysics, 60 Garden Street, Cambridge, Massachusetts 02138 USA.
[2]Niels Bohr Institute, Copenhagen University, Juliane Maries Vej 30, DK-2100 Copenhagen, Denmark.
[3]Observatoire de Genève, Université de Genève, 51 chemin des Maillettes, 1290 Sauverny, Switzerland. [4]Université Joseph Fourier – Grenoble 1, Centre national de la recherche scientifique, Laboratoire d'Astrophysique de Grenoble (LAOG), UMR 5571, 38041 Grenoble Cedex 09, France.
[5]Department of Physics, Kavli Institute for Astrophysics and Space Research, Massachusetts Institute of Technology, Cambridge, Massachusetts 02139, USA.

**A decade ago, the detection of the first[1,2] transiting extrasolar planet provided a direct constraint on its composition and opened the door to spectroscopic investigations of extrasolar planetary atmospheres[3]. As such characterization studies are feasible only for transiting systems that are both nearby and for which the planet-to-star radius ratio is relatively large, nearby small stars have been surveyed intensively. Doppler studies[4–6] and microlensing[7] have uncovered a population of planets with minimum masses of 1.9–10 times the Earth's mass ($M_\oplus$), called super-Earths. The first constraint on the bulk composition of this novel class of planets was afforded by CoRoT-7b (refs 8, 9), but the distance and size of its star preclude atmospheric studies in the foreseeable future. Here we report observations of the transiting planet GJ 1214b, which has a mass of 6.55$M_\oplus$ and a radius 2.68 times Earth's radius ($R_\oplus$), indicating that it is intermediate in stature between Earth and the ice giants of the Solar System. We find that the planetary mass and radius are consistent with a composition of primarily water enshrouded by a hydrogen–helium envelope that is only 0.05% of the mass of the planet. The atmosphere is probably escaping hydrodynamically, indicating that it has undergone significant evolution during its history. As the star is small and only 13 parsecs away, the planetary atmosphere is amenable to study with current observatories.**

The recently commissioned MEarth Project[10,11] uses an array of eight identical 40-cm automated telescopes to photometrically monitor 2,000 nearby M dwarfs with masses between 0.10 and 0.35 solar masses ($M_\odot$) drawn from a sample[12] of nearby stars with a large proper motion. After applying a trend-filtering algorithm[13] and a three-day running median filter to remove long-term stellar variability, we searched[14] the light curves for evidence of periodic eclipsing signals. The light curve of the star GJ 1214 contained 225 data points, of which six values were consistent with having been gathered during a time of eclipse and indicating a signal with a period of 1.58 days. On the basis of this prediction, we gathered additional photometric observations at high cadence using the eight telescopes of the MEarth array as well as the adjacent 1.2-m telescope. These light curves (shown in Fig. 1) confirm that the star is undergoing flat-bottomed eclipses with a depth of 1.3%, indicative of a planetary transit. Astrophysical false positives that result from blends of eclipsing binary stars and hinder field transit surveys are not[10,11] a concern under the strategy of the MEarth survey. GJ 1214 has a large proper motion, and by examining archival images we established that no second star lies at the current sky position of GJ 1214, ruling out a blend resulting from an eclipsing binary that is not physically associated with the target. The measured parallax and photometry of GJ 1214 (Table 1) place stringent constraints on the presence of an unresolved physically associated binary companion: we find no physically plausible coeval model that matches both the observed transit depth and the short duration of ingress and egress. We subsequently used the HARPS[5,6] instrument to gather radial velocity observations (Fig. 2 and Supplementary Information), which confirmed the planetary nature of the companion and permitted us to estimate its mass.

Table 1 presents our estimates of the physical quantities for planet and star. We estimate the planetary equilibrium temperature to be as great as 555 K (the case for a Bond albedo of 0) and as low as 393 K (assuming a Bond albedo of 0.75, the same as that for Venus). This latter value is significantly cooler than all known transiting planets, and exceeds the condensation point of water by only 20 K. This consideration is significant, because it demonstrates that for M dwarfs the discovery of super-Earths within the stellar habitable zones is within reach of ground-based observatories such as MEarth, whereas the discovery of such objects orbiting solar analogues is thought to require space-based platforms such as the Kepler Mission[15].

We compare in Fig. 3 the measured mass and radius of GJ 1214b with that of models[16] that predict planetary radii as a function of mass and assumed composition. We consider a hypothetical[16] water-dominated composition (75% $H_2O$, 22% Si and 3% Fe) and take this prediction to be an upper bound on the planet radius, assuming a solid composition. This model provides a minimum mass for the gaseous envelope: assuming that the envelope is isothermal (with a temperature corresponding to a Bond albedo of 0, above) and composed of hydrogen and helium, and that the observed planetary transit radius corresponds[17] to an atmospheric pressure of 1 mbar, we estimate a scale height of 233 km and a total envelope mass of $0.0032 M_\oplus$ (0.05% of the planetary mass). In this model, the relative mass of the envelope to the core is much smaller than that for the ice giants of the Solar System. If we continue under this assumed composition and consider both the Solar System planets and the extrasolar worlds together in Fig. 3, the sequence decreasing in mass from HD 149026b and Saturn to HAT-P-11b, GJ 436b, Neptune and Uranus, and finally GJ 1214b would then trace an atmospheric depletion curve: the mass of the gaseous envelope relative to that of the core would decrease with mass, which is consistent with the fact that the atmospheres of Earth and Venus are each only a trace component by mass. We note, however, that with only an estimate of the average density, we cannot be certain that GJ 1214b, GJ 436b and HAT-P-11b do not have compositions significantly different from that assumed above. For example, these planets could contain cores of iron or silicates enshrouded by much more massive envelopes of hydrogen and helium, a situation that would challenge models of formation but is not excluded by the current observations.

Our estimate of the stellar radius is 15% larger than that predicted by theoretical models[18] for the stellar mass we derived. Such discrepancies are well established from observations of M-dwarf eclipsing binaries, and indeed a similar stellar radius enhancement was determined[19] for the only other M-dwarf with a known transiting planet, GJ 436. If the true value of the stellar radius is $0.18 R_\odot$ (as predicted by both the theoretical models[18] and an empirical radius relation[20] for low-mass stars), then the planet radius would be revised downwards to $2.27 R_\oplus$, which is consistent with a water-dominated composition without the need for a gaseous envelope. If the empirical relation[21] for angular diameter can be extended to this spectral type, this would provide an alternative estimate of the stellar radius, given a refined estimate of the parallax.

We considered the timescale for hydrodynamic escape of a hydrogen-dominated envelope. Assuming that the ultraviolet luminosity of the star is $10^{-5}$ of its bolometric luminosity (typical[22] for inactive field M dwarfs), we calculate[23] a hydrodynamical escape rate of $9 \times 10^5$ kg s$^{-1}$; we further verified that at the sonic point the mean free path is only 4% of the scale height. At this rate, the minimum-mass envelope described above would be removed in about 700 Myr. The stellar ultraviolet radiation was probably much larger when the star was young, which would result in an even shorter timescale for removal of the envelope. An age of 3–10 Gyr for the star is supported[24] both by its kinematics (which indicate that it is a member of the old disk) and the lack of chromospheric activity from the absence of Hα line emission. Moreover, the dominant periodicity in the MEarth photometry is 83 days. Stars spin down as they age, and a very long rotation would also indicate an old star. Thus we conclude that significant loss of atmospheric mass has occurred over the lifetime of the planet; the current envelope is therefore probably not primordial. Moreover, some (or all) of the present envelope may have resulted from outgassing and further photodissociation of material from the core. If the composition of the gaseous envelope is indeed dominated by hydrogen (whether primordial or not), the annulus of the transmissive portion of planetary atmosphere would occult roughly 0.16% of the stellar disk during transit and thus present a signal larger than that already studied for other exoplanets[3]. Thus GJ 1214b presents an opportunity to study a non-primordial atmosphere enshrouding a world orbiting another star. Such studies have been awaited[25] and would serve to confirm directly that the atmosphere was predominantly hydrogen, because only then would the scale height be large enough to present a measurable wavelength-dependent signal in transit.

The discussion above assumes that the solid core of GJ 1214b is predominantly water. This is at odds with the recently discovered[8,9] CoRoT-7b, the only other known transiting super-Earth. CoRot-7b has mass of $4.8M_\oplus$, a radius of $1.7R_\oplus$ and a density of 5,600 kg m$^{-3}$, indicating a composition that is predominantly rock. The very different radii of GJ 1214b and CoRoT-7b despite their indistinguishable masses may be related to the differing degrees to which the two planets are irradiated by their parent stars: owing to the much greater luminosity of its central star, CoRoT-7b has an equilibrium temperature of about 2,000 K, roughly fourfold that of GJ 1214b. It may be that both planets have rocky cores of similar mass and that it is only for CoRoT-7b that the gaseous envelope has been removed, yielding the smaller observed radius. Alternatively, GJ 1214b may have a water-dominated core, indicating a very different formation history from that of CoRoT-7b. Such degeneracies in the models[16] of the physical structures of super-Earths will be commonplace when only a radius and mass are available, but at least one method[25] has been proposed to mitigate this problem in part. The differences in composition between GJ 1214b and CoRoT-7b bear on the quest for habitable worlds: numerous planets with masses indistinguishable from those of GJ 1214b and CoRoT-7b have been uncovered indirectly by radial velocity studies, and some of these lie in or near their stellar habitable zones. If such cooler super-Earth planets do indeed have gaseous envelopes similar to that of GJ 1214b, the extreme atmospheric pressure and absence of stellar radiation at the surface might render them inhospitable to life as we know it on Earth. This would motivate the push to even more sensitive ground-based techniques capable of detecting planets with sizes and masses equal to that of the Earth orbiting within the habitable zones of low-mass stars.

**Supplementary Information** is linked to the online version of the paper at www.nature.com/nature.

**Acknowledgements** We thank M. Everett for gathering the FLWO 1.2-m observations, S. Seager for providing a digital version of the structural models, and D. Sasselov and S. Seager for comments on the manuscript. Support for this work was provided by the David and Lucile Packard Foundation Fellowship for Science and Engineering awarded to D.C., and by the US National Science Foundation under grant number AST-0807690. L.B. and D.W.L. acknowledge support from the NASA Kepler mission under cooperative agreement NCC2-1390. M.J.H. acknowledges support by NASA Origins Grant NNX09AB33G. The HARPS observations were gathered under the European Southern Observatory Director's Discretionary Program 283.C-5022 (A). We thank the Smithsonian Astrophysical Observatory for supporting the MEarth Project at FLWO.

**Author Contributions** D.C., Z.B., J.I., C.B., P.N. and E.F. gathered and analysed the photometric data from the MEarth observatory, C.L., X.B., L.B., S.U., D.Q., F.P., M.M. and C.B. gathered and analysed the spectroscopic data from the HARPS instrument, and L.B., D.L., M.H., J.W. and P.N. gathered and analysed supplemental photometric and spectroscopic data with the 1.2-m and 1.5-m FLWO telescopes. R.M.-C. estimated the hydrodynamic escape rate, and X.B., X.D., T.F., J.I. and P.N. estimated the properties of the parent star. All authors discussed the results and commented on the manuscript. D.C. led the project and wrote the paper.

**Author Information** Reprints and permissions information is available at www.nature.com/reprints. The authors declare no competing financial interests. Correspondence and requests for materials should be addressed to D.C. (dcharbonneau@cfa.harvard.edu).

**Figure 1** | **Photometric data for GJ 1214.** Light curves of GJ 1214 spanning times of transit for four separate transit events, gathered with the MEarth Observatory (either a single telescope or eight telescopes, denoted respectively as MEarth × 1 and MEarth × 8) and the F. L. Whipple Observatory (FLWO) 1.2-m telescope. All light curves have been binned to a uniform cadence of 45 s to facilitate a visual comparison. We fitted the unbinned light curves to a model[29] corresponding to a planet in a circular orbit transiting a limb-darkened star, setting the limb-darkening coefficients to match the inferred stellar properties as described in the text. This model has five parameters: the orbital period $P$, the time of transit centre $T_c$, the ratio of the radius of the planet to that of the star $R_p/R_s$, the ratio of the semimajor axis to the stellar radius $a/R_s$, and the orbital inclination $i$. We use a Markov chain Monte Carlo method to estimate the uncertainties, and our results are stated in Table 1. The solid lines show the best-fit model fitted simultaneously to all the data.

**Figure 2** | **Change in radial velocity of GJ 1214. a**, We gathered 21 observations during 2009 July 24 to 2009 August 6, and six observations during 2009 June 11–19. We estimate[30] the change in the radial velocity by first constructing a stellar template by summing the observations (corrected to the barycentre), and then minimizing the $\chi^2$ difference between this template and each spectrum. We initially restricted our analysis to the July–August data (shown as filled points, with repetitions shown as open symbols), out of concern that long-term stellar variability or a second planet could lead to an offset between these data and those gathered in June (not shown). We fitted a sinusoidal model (solid curve) constrained by the photometric period and time of transit (dotted lines) and found a good fit ($\chi^2 = 15.98$ for 19 degrees of freedom) with a semi-amplitude of $K = 12.2 \pm 1.6$ m s$^{-1}$. We considered an eccentric orbit, and found that the best-fit model ($\chi^2 = 13.02$ for 17 degrees of freedom) was not significantly better and yielded an indistinguishable $K$. We conclude that there is no evidence that the orbit is non-circular, and we state the upper limit in Table 1. We then included the June observations and found

$K = 12.4 \pm 1.8$ m s$^{-1}$, which is consistent with but noisier than the previous estimate. However, to obtain a $\chi^2$ consistent with an acceptable fit, we need to introduce an additional noise term of 2.7 m s$^{-1}$, or an offset of $-8$ m s$^{-1}$ from the June data to the July–August data. Our photometry indicates that the stellar brightness varies by 2% on timescales of several weeks. We conclude that spot-induced stellar jitter is the most likely explanation. **b**, Residuals of the July–August data to the sinusoidal model. The residuals are consistent with the internal estimates of the uncertainties, shown here as $1\sigma$ error bars.

**Figure 3 | Masses and radii of transiting planets.** GJ 1214b is shown as a red filled circle (the $1\sigma$ uncertainties correspond to the size of the symbol), and the other known transiting planets are shown as open red circles. The eight planets of the Solar System are shown as black diamonds. GJ 1214b and CoRoT-7b are the only extrasolar planets with both well-determined masses and radii for which the values are less than those for the ice giants of the Solar System. Despite their indistinguishable masses, these two planets probably have very different compositions. Predicted[16] radii as a function of mass are shown for assumed compositions of H/He (solid line), pure H$_2$O (dashed line), a hypothetical[16] water-dominated world (75% H$_2$O, 22% Si and 3% Fe core; dotted line) and Earth-like (67.5% Si mantle and a 32.5% Fe core; dot-dashed line). The radius of GJ 1214b lies $0.49 \pm 0.13$ $R_\oplus$ above the water-world curve, indicating that even if the planet is predominantly water in composition, it probably has a substantial gaseous envelope.

**Table 1 | System parameters for GJ 1214**

| Parameter | Value |
|---|---|
| Orbital period, $P$ (days) | $1.5803925 \pm 0.0000117$ |
| Times of centre of transit, $T_c$ (HJD) | $2454964.944208 \pm 0.000403$ |
| | $2454980.7479702 \pm 0.0000903$ |
| | $2454983.9087558 \pm 0.0000901$ |
| | $2454999.712703 \pm 0.000126$ |
| Planet/star radius ratio, $R_p/R_s$ | $0.1162 \pm 0.00067$ |
| Scaled semimajor axis, $a/R_s$ | $14.66 \pm 0.41$ |
| Impact parameter, $b$ | $0.354^{+0.061}_{-0.082}$ |
| Orbital inclination, $i$ (deg) | $88.62^{+0.35}_{-0.28}$ |
| Radial velocity semi-amplitude, $K$ (m s$^{-1}$) | $12.2 \pm 1.6$ |
| Systemic velocity, $\gamma$ (m s$^{-1}$) | $-21{,}100 \pm 1{,}000$ |
| Orbital eccentricity, $e$ | <0.27 (95% confidence) |
| Stellar mass, $M_s$ | $0.157 \pm 0.019 M_\odot$ |
| Stellar radius, $R_s$ | $0.2110 \pm 0.0097 R_\odot$ |
| Stellar density, $\rho_s$ (kg m$^{-3}$) | $23{,}900 \pm 2{,}100$ |
| Log of stellar surface gravity (CGS units), $\log g_s$ | $4.991 \pm 0.029$ |
| Stellar projected rotational velocity, $v \sin i$ (km s$^{-1}$) | <2.0 |
| Stellar parallax (mas) | $77.2 \pm 5.4$ |
| Stellar photometry | |
| $V$ | $15.1 \pm 0.6$ |
| $I$ | $11.52 \pm 0.1$ |
| $J$ | $9.750 \pm 0.024$ |
| $H$ | $9.094 \pm 0.024$ |
| $K$ | $8.782 \pm 0.020$ |
| Stellar luminosity, $L_s$ | $0.00328 \pm 0.00045 L_\odot$ |
| Stellar effective temperature, $T_{eff}$ (K) | $3{,}026 \pm 130$ |
| Planetary radius, $R_p$ | $2.678 \pm 0.13 R_\oplus$ |
| Planetary mass, $M_p$ | $6.55 \pm 0.98 M_\oplus$ |
| Planetary density, $\rho_p$ (kg m$^{-3}$) | $1870 \pm 400$ |
| Planetary surface acceleration under gravity, $g_p$ (m s$^{-2}$) | $8.93 \pm 1.3$ |
| Planetary equilibrium temperature, $T_{eq}$ (K) | |
|     Assuming a Bond albedo of 0 | 555 |
|     Assuming a Bond albedo of 0.75 | 393 |

To convert the photometric and radial velocity parameters into physical parameters for the system, we require a constraint on the stellar mass. Using the observed parallax distance[26] of $12.95 \pm 0.9$ pc and apparent $K$-band brightness, we employ an empirical relation[27] between stellar mass and absolute $K$-band magnitude to estimate the stellar mass. With this value we find the planetary radius and mass. The uncertainty on the planet mass is the quadrature sum of the propagated uncertainties on the radial-velocity amplitude and those from the uncertainty in the stellar mass, which contribute $0.85 M_\oplus$ and $0.50 M_\oplus$ to the error budget, respectively. We use the observed $I - K$ colour and an empirical relation[28] to estimate the bolometric correction and subsequently the stellar luminosity and stellar effective temperature (assuming the stellar radius quoted in the table). Using the luminosity, we estimate a planetary equilibrium temperature, assuming a value for the Bond albedo. HJD, heliocentric Julian date.

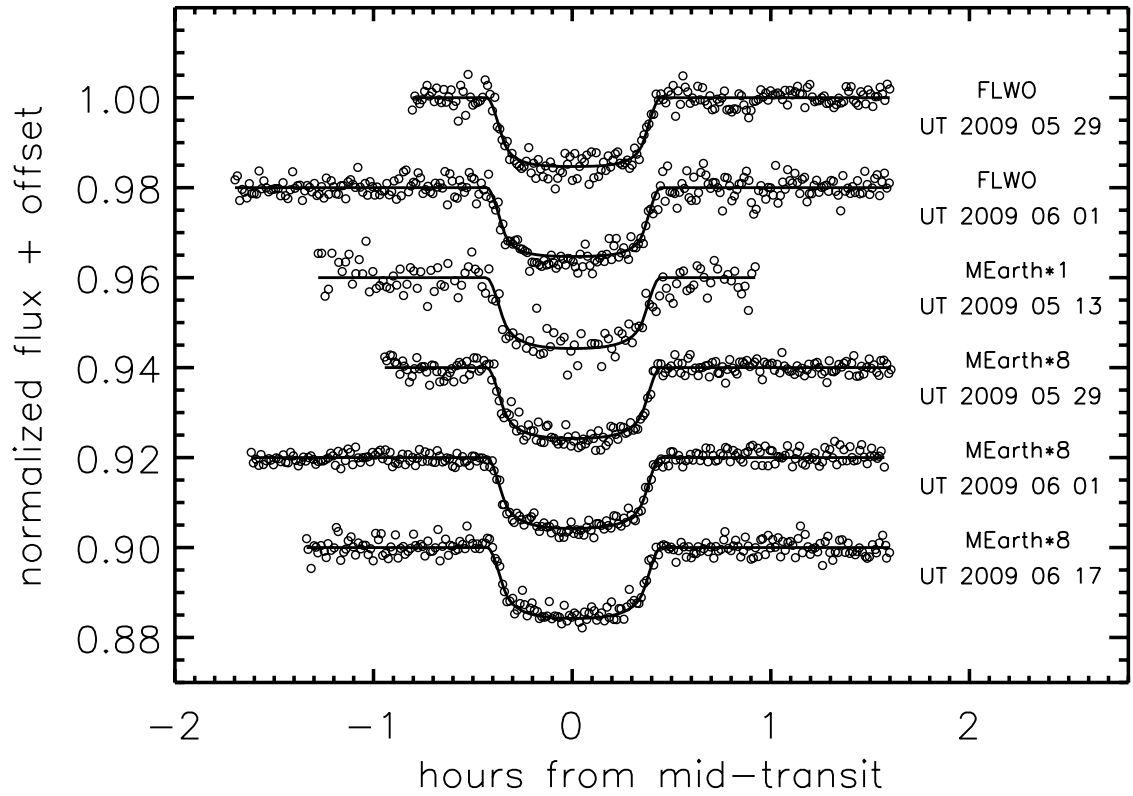

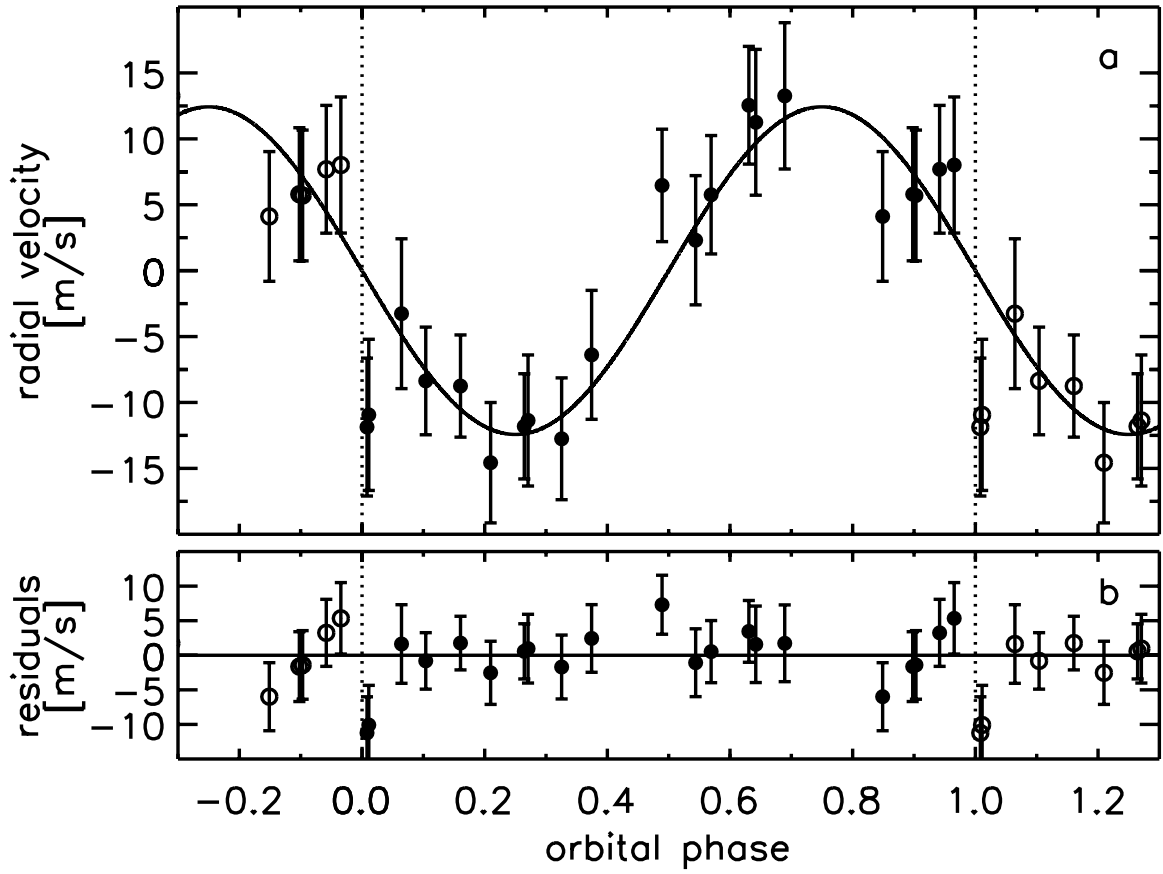

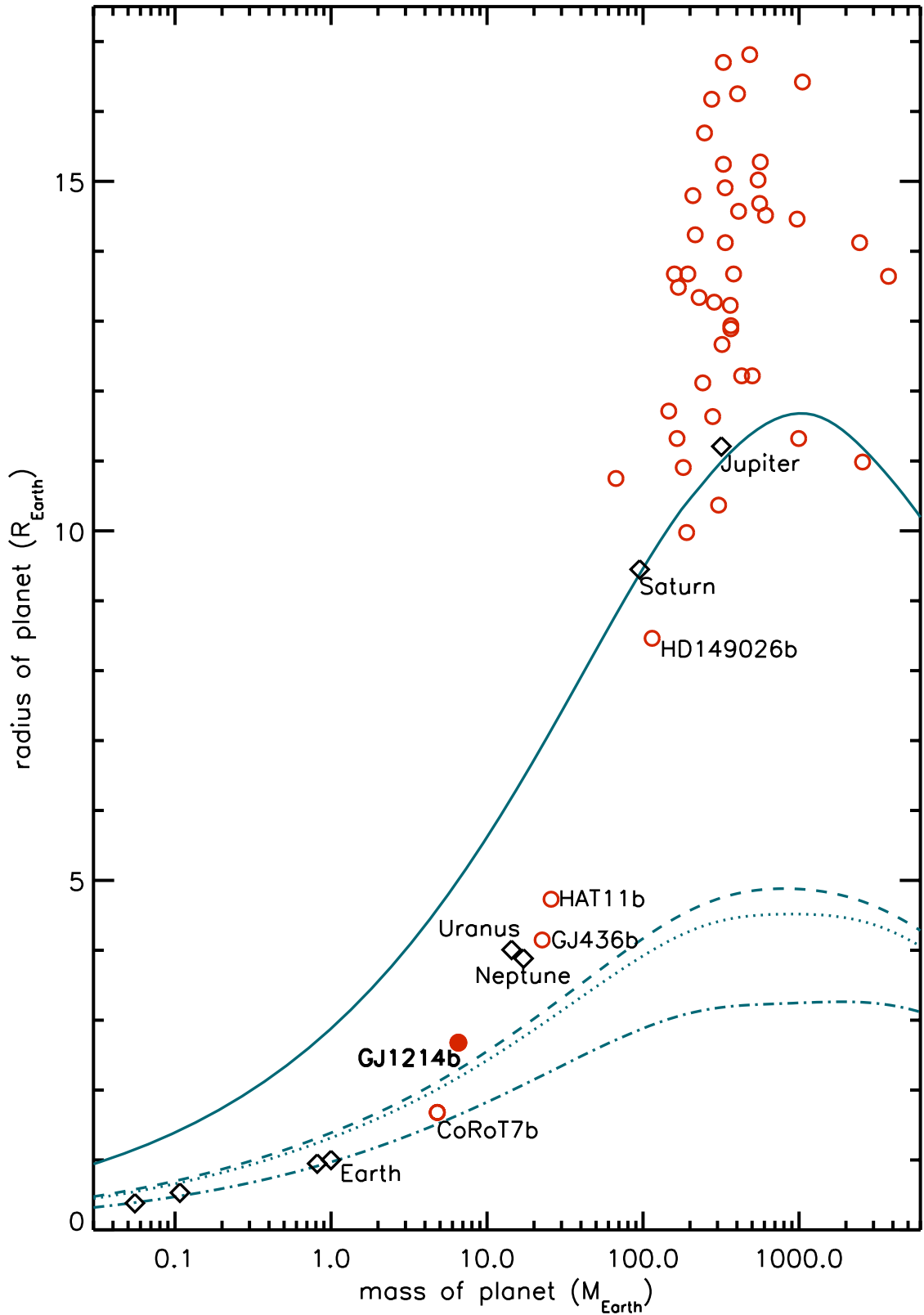